%% file: main.tex
\documentclass[12pt]{article}
\title{}
\author{}
\date{}
\usepackage[utf8]{inputenc}
\usepackage[letterpaper,top=1in,right=1in,left=1in,bottom=1in]{geometry}
\usepackage{amsmath}
\usepackage{amsfonts}
\usepackage{amssymb}
\usepackage{graphics}
\usepackage{graphicx}
\usepackage{placeins}
\usepackage[capposition=top]{floatrow}
\usepackage{subcaption}
\usepackage{lscape}
\usepackage{pdflscape}
\usepackage{setspace}
\usepackage{standalone}
\usepackage{fancyhdr}
\usepackage{framed}
\usepackage{color}
\usepackage{tikz}
\usepackage{pgfplots}
\usepackage{longtable}
\usepackage{booktabs,caption}
\usepackage[draft]{todonotes}
\usepackage{booktabs,caption}
\usepackage[flushleft]{threeparttable}
\usepackage{bbm}
\usepackage{multirow}
\usepackage{lineno}
\usepackage{titlesec}
\usepackage{hyperref}
\usepackage{comment}
\usepackage[draft]{todonotes}
\usepackage[justification=centering]{caption}
\pgfplotsset{compat=1.15}

\usepackage{longtable}
\usepackage{array}
\usepackage{authblk}
\usepackage{pdflscape}
\usepackage{colortbl}
\usepackage{ragged2e}
\usepackage{xfp}
\usepackage{lipsum}

\usepackage{stackengine}
\usepackage{array}
\newcolumntype{L}[1]{>{\raggedright\arraybackslash}p{#1}}
\setstackEOL{\#}
\setstackgap{L}{12pt}

\usepackage[page,toc,titletoc,title]{appendix}
\usepackage{blindtext}
\usepackage{pdfpages}

\hypersetup{
	colorlinks=true, 
	linktoc=all,     
	linkcolor=blue,  
	citecolor=blue,
    urlcolor=blue
}

\usepackage[style=apa,natbib]{biblatex}
\usepackage[all]{nowidow} 
\usepackage[protrusion=true,expansion=true]{microtype} 
\usepackage{orcidlink}

\hypersetup{ 	
pdfsubject = {},
pdftitle = {2PL EIRM},
pdfauthor = {Gilbert, Joshua B.}
}

\usepackage{setspace}

\title{Polytomous Explanatory Item Response Models for Item Discrimination: Assessing Negative-Framing Effects in Social-Emotional Learning Surveys}

\begin{document}

\begin{titlepage}
\author[1]{Joshua B. Gilbert\,\orcidlink{0000-0003-3496-2710}}
\author[2]{Lijin Zhang\,\orcidlink{0000-0002-4222-8850}}
\author[3,4]{Esther Ulitzsch\, \orcidlink{0000-0002-9267-8542}}
\author[2]{Benjamin W. Domingue\,\orcidlink{0000-0002-3894-9049}}

\affil[1]{Harvard University Graduate School of Education}
\affil[2]{Stanford University Graduate School of Education}
\affil[3]{Centre for Educational Measurement (CEMO), University of Oslo}
\affil[4]{Centre for Research on Equality in Education (CREATE), University of Oslo}

\maketitle

\begin{abstract}

\noindent Modeling item parameters as a function of item characteristics has a long history but has generally focused on models for item location. Explanatory item response models for item discrimination are available but rarely used. In this study, we extend existing approaches for modeling item discrimination from dichotomous to polytomous item responses. We illustrate our proposed approach with an application to four social-emotional learning surveys of preschool children to investigate how item discrimination depends on whether an item is positively or negatively framed. Negative framing predicts significantly lower item discrimination on two of the four surveys, and a plausibly causal estimate from a regression discontinuity analysis shows that negative framing reduces discrimination by about 30\% on one survey. We conclude with a discussion of potential applications of explanatory models for item discrimination. \\


Corresponding author: \href{mailto:joshua_gilbert@g.harvard.edu}{joshua\_gilbert@g.harvard.edu}

\end{abstract}
\end{titlepage}

\doublespacing

\section{Introduction} \label{intro}

Why are some items easier, harder, or more discriminating than others? This question has posed a longstanding challenge in psychometrics and Item Response Theory (IRT). From the linear logistic test model \citep{fischer1973linear} to explanatory item response models \citep{wilson2008explanatory, de2004explanatory, de2016explanatory}, causal Rasch models \citep{stenner2013causal}, and contemporary machine learning approaches \citep{vstvepanek2023item, alkhuzaey2023text, pandarova2019predicting}, various methods attempt to model item parameters as functions of item features. The appeal of such models lies in the potential that test makers can design more reliable and informative measures if they can understand why items function the way they do.

Extensive literature on modeling item parameters has focused on explaining differences in item location (i.e., difficulty, easiness, agreeability) as a function of item characteristics, but the attempt to explain differences in item discrimination is comparatively rare \citep{cho2014additive, petscher2020past,embretson1999generating}. Understanding why some items are more discriminating than others could lead to more informative and efficient test design because more discriminating items provide more information about respondents whose latent trait values are near the item location parameters \citep{van2017handbook1, an2014item}. Furthermore, systematic differences in item discrimination could also shed light on fundamental measurement issues such as scaling, differential item functioning, and the cognitive processes that underlie item responses \citep{tuerlinckx2005two, humphry2011role, masters1988item, jordan2019rethinking}. 

Early model developments for relating item discrimination to item characteristics mirror the linear logistic test model (LLTM) in that they assume that variation in item discrimination can be perfectly explained by item characteristics \citep{embretson1999generating}. To relax this assumption, subsequent approaches include error terms that capture the variability in item discrimination that is not explained by the item characteristics and thus provide a more realistic modeling framework \citep{cho2014additive, klein2009evaluating}. Despite their potential utility, however, empirical applications of these models appear to be quite rare due to the specialized software required for model estimation, with very few extant examples in the published literature that we briefly review here. 

\textcite{cho2014additive}  examine mathematics items from the Graduate Record Examination (GRE) and find that estimated effects of item characteristics on item discrimination are weak (p. 98). Somewhat similar to the present study, \textcite{min2018understanding} examine item wording effects on polytomous items from the Big Five personality scale, and find that negative wording was positively associated with item discrimination for four out of five subscales. However, the relationships between item characteristics and item discrimination parameters lack standard errors due to software constraints (p. 3) and therefore the results are inconclusive and suggest the need for alternative modeling approaches. \textcite{petscher2020past} explore applications in elementary school literacy and dyslexia and show that word decodability and valence positively and negatively explain item discrimination, respectively. Last, \textcite{zorowitz2024item} examine matrix reasoning tasks and find that the item characteristic  rule number (``the number of relationships that govern the changes among these shapes across cells'', pp. 1108-1109) significantly predicts item discrimination.\footnote{Previous applications of random effects loading models in the factor analysis and SEM traditions have emphasized the use of random factor loadings to handle large numbers of items or multilevel structures, rather than explanatory models for the loadings themselves \citep{muthen2018recent, asparouhov2012general}. }

In this study, we pursue two goals. First, we review the conventional explanatory item response model (EIRM) for item location and show how the model can be extended to allow for varying item discriminations and estimated with open-source software. Second, we extend previous work on explanatory models for item discrimination to accommodate polytomous item responses and a multiplicative structure for the item discriminations (e.g. \cite{bolsinova2017modelling}). We therefore build on existing models for dichotomous responses that assume an additive discrimination structure \citep{cho2014additive,klein2009evaluating}. Thus, we synthesize the two-parameter logistic (2PL) approach of \textcite{cho2014additive} for dichotomous responses with subsequent applications of the 1PL EIRM to polytomous data \citep{huang2023explanatory, kim2020polytomous, kim_wilson2020polytomous, aydin2023modelling}. Our approach therefore extends prior frameworks by allowing for \textit{both} polytomous responses and varying item discriminations with item random effects and makes models for item discrimination more accessible and applicable to researchers in a variety of empirical settings such as education, psychology, and behavioral research. 

We motivate and illustrate our proposed approach with an empirical application to four psychological survey scales measuring various dimensions of social-emotional learning (SEL) in preschool children. Our primary research question is to determine whether negatively-framed items (i.e., items that need to be reverse-coded because a higher score represents a lower level of the latent trait) are systematically less discriminating than positively-framed items. Understanding systematic differences in discrimination between positively- and negatively-framed items could provide useful information for the development of measures and has been the subject of considerable discussion in the measurement literature (e.g. \cite{weems2001impact, venta2022reverse, suarez2018using, weijters2012misresponse, sliter2014irt, greenberger2003item, barker1982comparison, kam2018we}). For example, if negatively-framed items provide consistently lower discrimination than positively-framed items due to, say, increased cognitive load, an argument could be made for using them more sparingly to maximize the efficiency of measurement. Our approach therefore complements other research that examines item wording and framing effects from the perspective of multidimensionality and bi-factor models in the structural equation modeling and factor analysis tradition \citep{gu2015impact, tang2024examination, horan2003wording}. Our results show that in two of the four surveys, the negatively-framed items were substantially less discriminating than the positively-framed items. Furthermore, a plausibly causal estimate from a regression discontinuity analysis in one survey suggests that negative framing reduces item discrimination by about 30\%.

\section{The Explanatory Item Response Model (EIRM)} \label{eirm}

Originally formalized by \textcite{de2004explanatory}, the EIRM builds on previous work that synthesizes IRT and multilevel models into a unified framework \citep{rijmen2003nonlinear, adams1997multidimensional}. Consider the following model for dichotomous item responses with random effects for both persons and items \citep{de2008random, holland1990sampling}:

\begin{align}
\label{eq:eirm_desc}
    \text{logit}(\Pr(y_{ij} = 1)) = \eta_{ij} &= \theta_j + b_i \\
    \label{eq:b_i}
    b_i &= \beta_0 + \zeta_{0i} \\
    \theta_j  &\sim N(0, \sigma^2_\theta) \\
    \zeta_{0i} &\sim N(0, \sigma^2_b),
\end{align}

\noindent where $y_{ij}$ is the observed item response to item $i$ by person $j$,  $\theta_j$ is latent person ability, and $b_i$ is item location, commonly denoted item easiness on a cognitive measure or agreeability on an affective survey. Item location $b_i$ is in turn a function of the mean $\beta_0$ and item-specific deviation $\zeta_{0i}$. Without predictors in the model, Equation \ref{eq:eirm_desc} is equivalent to a one-parameter logistic (1PL) or Rasch IRT model, but with random effects for the items instead of the more typical item fixed effects. Equation \ref{eq:eirm_desc} is a ``doubly descriptive'' model in the typology of \textcite{wilson2008explanatory}, because it estimates variances for persons and items with no predictors to explain mean differences in $\theta_j$ or $b_i$ as a function of person or item characteristics. The model becomes item explanatory, person explanatory, or doubly explanatory when item, person, or both types of predictors are added to the model, respectively.

We can extend Equation \ref{eq:b_i} to include some covariate $x_i$ in the equation for $b_i$ to determine the relationship between $x_i$ and item location:

\begin{equation}
    b_i = \beta_0 + \beta_1 x_i + \zeta_{0i}.
\end{equation}

\noindent In our empirical application, $x_i$ indicates whether an item is negatively or positively framed in a psychological survey; estimates would determine whether there are systematic differences in item location based on item framing. The item location residuals $\zeta_{0i}$ allow for imperfect prediction, in contrast to more restrictive models such as the LLTM that assume that the item characteristics can perfectly explain the item location parameters. At the person level, we could similarly include some covariate $x_j$ in the equation for $\theta_j$ to estimate mean differences in the latent trait as a function of $x_j$. Interactions between person and item predictors can capture uniform differential item functioning (DIF;  \cite{de2011estimation, randall2011using}). 

Given its flexibility, the EIRM has seen many applications in the psychometric, psychological, educational, and behavioral literature, including extensions such as longitudinal models, multidimensionality, descriptive comparisons of group performance, DIF analysis, causal inference, and treatment heterogeneity (e.g. \cite{briggs2008using, gilbert2023jebs, gilbert2024disentangling, gilbert2024estimating, gilbert2024ipd, de2014multidimensional, cho2013measuring, kim2021influences, gilbert2011word, goodwin2013morphological, gilbert2024econ}). Recent work has extended the EIRM to polytomous response models, such as the rating scale model and the partial credit model, with fixed items \citep{bulut2021estimating, stanke2019explanatory} or random items \citep{kim2020polytomous, gilbert2024ssri, huang2023explanatory}, further extending its applicability to diverse research and assessment contexts.

One limitation of the traditional EIRM approach is that most applications have used 1PL or Rasch models that assume that the discriminating power of each item is equal with respect to $\theta_j$, a restrictive assumption unlikely to be met in much empirical data \citep{mcneish2020thinking}. In part, this limitation is due to software constraints. Several widely cited papers demonstrate how to use the free and open-source multilevel modeling software \texttt{lme4} to fit the EIRM in R \citep{doran2007estimating, de2011estimation}, but \texttt{lme4} cannot accommodate varying discrimination parameters by item unless they are known in advance \citep{rockwood2019estimating}. Although proprietary or specialized software such as \texttt{Mplus} \citep{Muthen2017}, and Stata's \texttt{gllamm} \citep{rabe2003maximum} can accommodate the 2PL EIRM with item random effects, empirical applications have been rare \citep{petscher2020past}.

Recently, three open-source software programs in R use novel estimation methods that allow for the estimation of item discrimination parameters in the EIRM, namely, \texttt{PLmixed} \citep{rockwood2019estimating}, \texttt{galamm} \citep{sorensen2024multilevel}, and \texttt{brms} \citep{burkner2021brms}. While \texttt{PLmixed} and \texttt{galamm} represent an advance in the flexibility of models that can be fit, they do not allow for random item discriminations or polytomous responses, thus limiting their utility when the goal is to explain \textit{differences} in item discrimination as a function of covariates. In contrast, \texttt{brms} provides the most flexible approach that allows for a wide range of IRT models with fixed or random locations, discriminations, and link functions, providing the most attractive option for explanatory modeling of item discriminations under investigation in this study. We include further discussion of the features and limitations of various alternative IRT software packages available in R in Appendix \ref{appendix:software}.

Given these advances, consider the addition of a discrimination parameter $a_i$ to the 2PL EIRM:

\begin{align}
   \label{eq:eirm_2pl}
    \eta_{ij} &= a_i(\theta_j + b_i) \\
    b_i &= \beta_0 + \zeta_{0i} \\
    \label{eq:a}
    \ln(a_i) &= \gamma_0 + \zeta_{1i} \\ 
   \theta_j &\sim N(0, 1) \\
    \begin{bmatrix}
        \zeta_{0i} \\
        \zeta_{1i}
    \end{bmatrix} &\sim 
    N\left(\begin{bmatrix}
        0 \\ 0
    \end{bmatrix},
    \begin{bmatrix}
        \sigma^2_b & \sigma_{ab} \\
        \sigma_{ab} & \sigma^2_a
    \end{bmatrix}\right).
\end{align}
\noindent In contrast to the 1PL or Rasch model in Equation \ref{eq:eirm_desc}, which uses the unit loading identification, here, $\sigma^2_\theta$ is constrained to 1 for model identification \citep{burkner2021brms}. The item discrimination parameter $a_i$ represents the degree to which item $i$ discriminates between respondents along the latent trait $\theta_j$. That is, a 1 standard deviation increase in $\theta_j$ causes an $a_i$ increase in the log-odds of a correct response for item $i$. $a_i$ and $b_i$ may be correlated, represented by the covariance parameter $\sigma_{ab}$.

An affordance of Equation \ref{eq:a} is that the $a_i$ are given a model in which $\gamma_0$ is the discrimination of the average item and $\zeta_{1i}$ is the deviation of item $i$ from the average with SD $\sigma_a$. While previous models are linear in the $a_i$ term \citep{cho2014additive, klein2009evaluating}, we follow \textcite{bolsinova2017modelling} and use a multiplicative or logarithmic link function for $a_i$. By modeling $a_i$ on the log scale, the $a_i$ are constrained to be positive, which provides benefits for the stability of model identification \citep[p. 26]{burkner2021brms}, and the residuals $\zeta_{1i}$ are assumed to follow a log-normal distribution, a common assumption in 2PL IRT modeling \citep[p. 88]{cho2014additive}. To our knowledge, applications of the model represented in Equation \ref{eq:eirm_2pl} and related approaches have thus far only considered dichotomous responses (e.g. \cite{petscher2020past, zorowitz2024item, cho2014additive}), thus limiting applicability in diverse assessment contexts where polytomous response data is common, such as educational, psychological, or behavioral research. We describe how to extend the model to polytomous responses in the context of our empirical application in Section \ref{methods}.

While traditionally modeled as fixed effects in IRT, a key benefit of the random effects approach to the $a_i$ is that we can include covariates in a latent regression to explain differences in item discrimination as a function of item or person characteristics. For example, the motivating application of this study includes an indicator variable for whether an item is negatively framed to determine if there are systematic differences in item discrimination by item framing. Applying the EIRM to address differences in item discrimination can also answer many other substantively important questions in educational, psychological, or behavioral measurement, to which we return in the discussion. 

\section{Methods} \label{methods}

To illustrate our proposed approach, we use publicly available item-level data from a study of social-emotional learning (SEL) among preschool children in the United States \citep{bailey2023sel}. The data set includes four common SEL measures for young children rated by adults at two time points, measuring emotional regulation \citep{shields1997emotion}, learning behaviors \citep{mcdermott2002development}, social competence \citep{lafreniere1996social}, and social skills \citep{gresham2008social}. The measures are summarized in Table \ref{table:measures} and include example positively- and negatively-framed items from each measure. All items are rated on a Likert scale and include identical response options within each survey, although each survey has a different number of response options. For the purposes of this study, we focus on the baseline assessments. (We include a parallel analysis of the followup assessment as a cross-validation of our empirical results in Appendix \ref{appendix:t2}.) The student sample contains 1000 children, who range in age from 3 to 5 years and are 49\% female.

Our empirical question is whether there are systematic differences in item discrimination between positively- and negatively-framed items. To illustrate, consider example items from the first measure, the Emotion Regulation Checklist (ERC). Here, a positively-framed item is ``is able to delay gratification,'' in that a higher score indicates greater emotional regulation. A negatively-framed item is ``exhibits wide mood swings,'' because a higher score indicates lower emotional regulation. With the exception of three items that included explicit negation (i.e., use of the word ``not''), all negatively-framed items were reverse worded such that a higher item score indicates a lower level of the latent trait, analogous to the preceding example \citep{sonderen2013ineffectiveness}. After completing a qualitative review of item text to identify negatively-framed items for each survey, we reverse-coded all item scores for the negatively-framed items so that a higher score indicates a greater degree of the latent trait. We include the full item text from each survey and our coding of the item framing in Appendix \ref{appendix:item_text}.

\begin{table}[ht]
\caption{Summary of SEL Measures}
\label{table:measures}
\centering
\input{tab1}
\justify \footnotesize
Notes:  The ERC typically contains 24 items, however, the replication materials from \textcite{bailey2023sel} indicate that question 7 was skipped in error. The SCBE contains six response categories but only four descriptive labels; i.e., ``sometimes'' could be rated as a 2 or 3. The data also included the Child Behavior Rating Scale, however, this 10-item measure contained no negatively-framed items so we exclude it from our analysis.
\end{table}

To motivate formal models for the item discrimination parameters, we begin with a descriptive analysis and fit separate graded response models (GRMs) with \texttt{mirt} \citep{chalmers2012mirt} to each survey and collect the $a_i$ estimates, with the added constraint that the distances between the category thresholds are equal across items (i.e., the rating scale framework; \cite{muraki1990fitting}), to match the explanatory models that we fit using \texttt{brms}. \texttt{brms} does not yet support order-preserving random effects models that allow for unique distances between thresholds by item \citep[p. 36]{burkner2021brms}, though the fact that all items share identical response options on each survey means that this constraint is less relevant compared to other cases with different rating scales by item. Furthermore, this constraint helps stabilize the estimates of category thresholds for item-category combinations with limited data and provides an interpretational benefit because each item has a single global location parameter \citep[pp. 103-104]{embretson2000irt}. 

To formally test differences in item discrimination by item framing, we extend the dichotomous response model of Equation \ref{eq:eirm_2pl} to estimate the following explanatory GRM, where $\text{negative}_i=1$ indicates that item $i$ is negatively framed and therefore reverse coded:
\begin{align}
\label{eq:egrm}
    \text{logit}(\Pr(y_{ij} \le k)) = \eta_{ijk} &= a_i(\alpha_k - (\theta_j + b_{i})) \\
    b_{i} &= \beta_1 \text{negative}_i + \zeta_{0i} \\
    \label{eq:ln_a}
    \ln(a_i) &= \gamma_0 + \gamma_1 \text{negative}_i + \zeta_{1i} \\
    \theta_j &\sim N(0, 1) \\
      \begin{bmatrix}
        \zeta_{0i} \\
        \zeta_{1i}
    \end{bmatrix} &\sim 
    N\left(\begin{bmatrix}
        0 \\ 0
    \end{bmatrix},
    \begin{bmatrix}
        \sigma^2_b & \sigma_{ab} \\
        \sigma_{ab} & \sigma^2_a
    \end{bmatrix}\right).
\end{align}

\noindent Here, we model the log-odds that person $j$ responds to item $i$ with a rating of less than or equal to $k$, where $k \in \{1,...,K-1\}$, $K$ is the total number of categories, $\Pr(y_{ij} \le 0) = 0$, and $\Pr(y_{ij} \le K = 1)$. $\alpha_k$ are fixed intercepts for scoring at or below category $k$ for the average positively-framed item, $\theta_j$ is the latent trait, and $b_{i}$ is an item-specific residual shift representing item agreeability ($\zeta_{0i}$) plus the average difference in agreeability between positively- and negatively-framed items ($\beta_1$). Thus, the distance between the intercepts is constant across items, but each item may be more or less agreeable than the average item. $\theta_j$ and $b_i$ are subtracted from $\alpha_k$ so that higher values predict higher response categories, following the standard parameterization of ordered logistic regression in most software \citep{bilder2014analysis, christensen2015analysis, ripley2013package, williams2006generalized}. The parameter of interest is $\gamma_1$, which captures differences in item discrimination between the item types on the log scale. A negative value of $\gamma_1$ would suggest that the negatively-framed items are less discriminating, on average, than positively-framed items. We fit the model separately to each of the four survey measures and apply weakly informative priors for the random effects, following the tutorial provided by \textcite{burkner2021brms}. Appendix \ref{appendix:code} shows example \texttt{brms} code to fit Equation \ref{eq:egrm} and associated models. Appendix \ref{appendix:dag} shows a directed acyclic graph (DAG) for Equation \ref{eq:egrm}.

In addition to the descriptive analyses offered by Equation \ref{eq:egrm}, the design of the Social Skills Improvement System (SSIS) measure allows for an additional causal analysis. Unlike the other measures, the SSIS includes all of the positively-framed items first and then switches to the negatively-framed items at item 47 out of 76. This design lends itself to a regression discontinuity (RD) analysis to estimate the causal effect of changing to negatively-framed items partway through the survey, a potentially more informative quantity than our other models that simply provide average differences in discrimination by item framing. We fit this model by extending Equation \ref{eq:ln_a} as follows in a linear RD framework:
 \begin{align}
 \label{eq:rd}
     \ln(a_i) &= \gamma_0 + \gamma_1 \text{negative}_i + \gamma_2 \text{position}_i + \gamma_3 \text{negative}_i \times \text{position}_i + \zeta_{1i},
 \end{align}
\noindent where $\text{position}_i$ is the position of item $i$ in the SSIS, centered at the boundary between the positive and negative items. 

The identifying assumption supporting a causal interpretation of $\gamma_1$ in Equation \ref{eq:rd} is that the potential item discriminations are smooth at the boundary between positively-framed and negatively-framed items. That is, the position of an item in the SSIS can affect its discrimination, for example through fatigue effects \citep{davis2005using}, but we would expect such effects to be a smooth function of item position. In other words, in the absence of the switch to negative framing, average item discrimination would be a smooth function of its position in the survey. Under this assumption, any discontinuity in discrimination observed at the boundary represented by $\gamma_1$ would provide evidence of a causal effect of the switch to negative framing on item discrimination partway through the SSIS \citep{hahn2001identification, angrist2009mostly, cunningham2021causal}.

\section{Results}

\subsection{Descriptive Analyses}

\begin{figure}
    \centering
    \includegraphics[width=.75\linewidth]{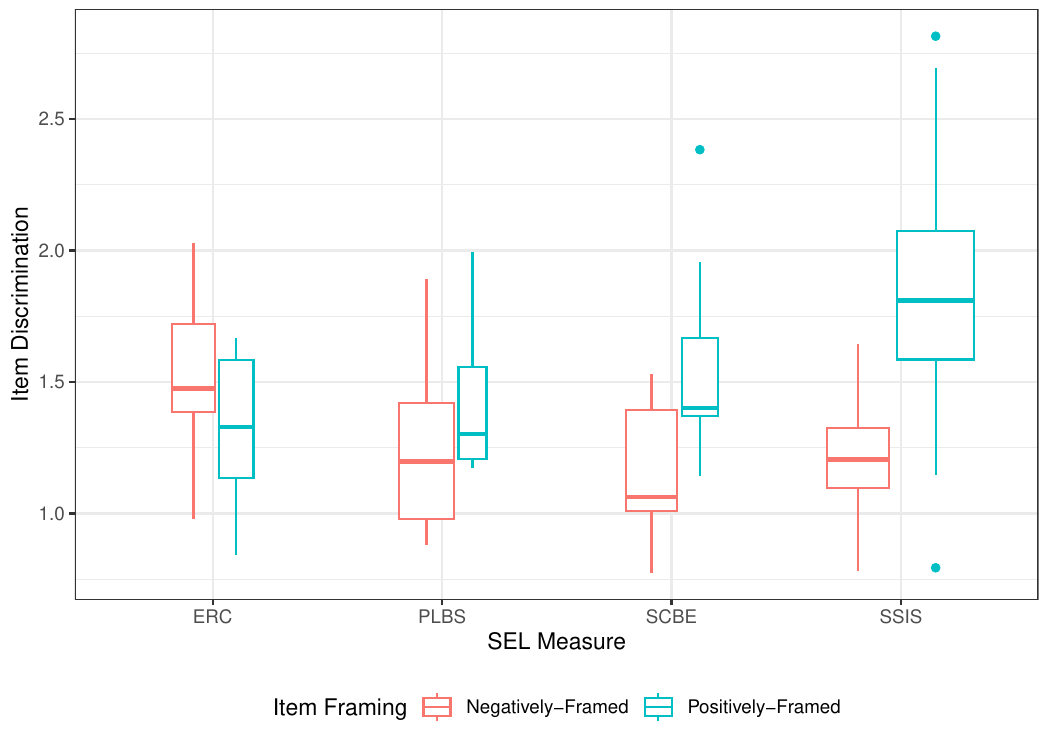}
    \caption{Boxplots of estimated $a_i$ from each survey}
    \label{fig:a_density}
                 \justify \footnotesize
Notes: ERC = emotion regulation checklist; PLBS = preschool learning behavior scale; SCBE = social competence and behavior evaluation; SSIS = social skills improvement system. The width of each box is proportional to the number of items. Boxes are color-coded by whether the item wording is positively or negatively framed (see Table \ref{table:measures} for example item text).
\end{figure}

The distributions of the estimated $a_i$ parameters from the GRMs calculated in \texttt{mirt} (without covariates) are shown in Figure \ref{fig:a_density}. Several facts merit attention. First, there appears to be a large range of $a_i$ in each survey, suggesting that the application of 1PL models is inappropriate. Second, all surveys except for the ERC show lower median discrimination parameters for negatively-framed items. Although descriptive analysis provides useful motivation, it suffers from important limitations. That is, it treats the $a_i$ as known when they are estimated with error, and, without a formal model, estimating the uncertainty or determining the significance of the apparent difference is challenging. While non-parametric approaches such as the bootstrap provide a reasonable alternative in the present case, bootstrapping methods would not be easily applicable to more complex models that consider combinations of person and item predictors. 

\begin{figure}
    \centering
    \includegraphics[width=.75\linewidth]{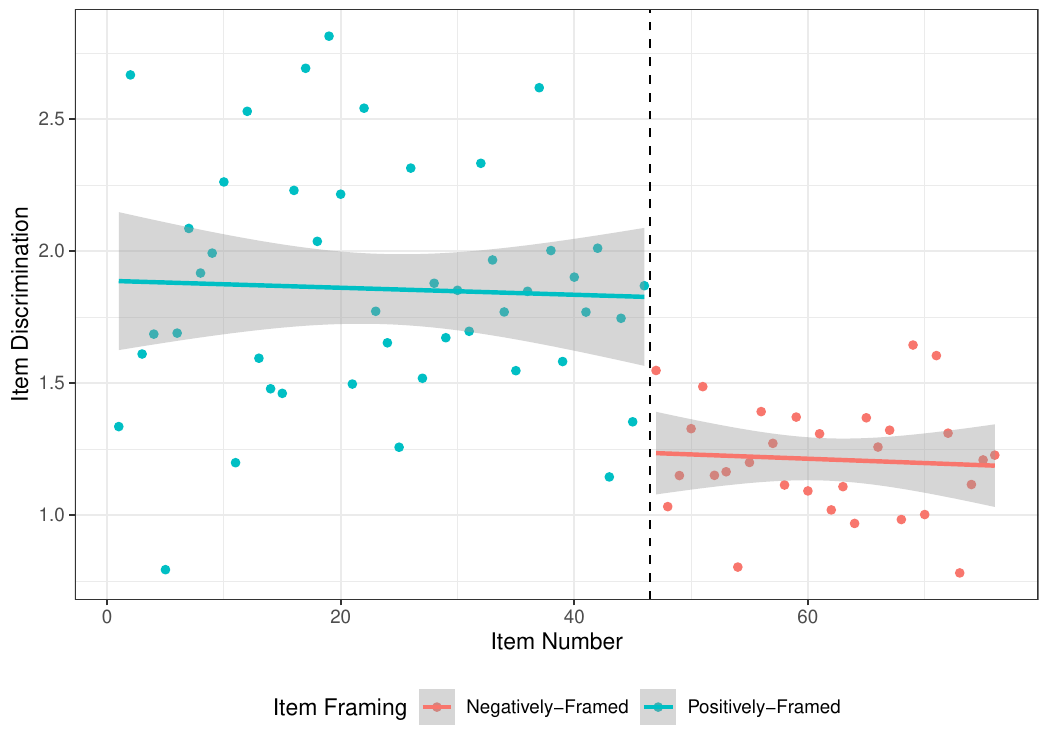}
    \caption{Estimated $a_i$ by item number for the SSIS}
    \label{fig:rd_descriptive}
                 \justify \footnotesize
Notes: SSIS = social skills improvement system. The lines and shaded areas represent OLS predictions and 95\% confidence intervals, respectively, fit separately to each item group. The points and lines are color-coded by whether the item wording is positively or negatively framed.
\end{figure}

For the RD analysis, Figure \ref{fig:rd_descriptive} shows a scatter plot of $a_i$ on item position on the SSIS and shows an apparent discontinuity at the cutoff, suggesting that the switch to negatively-framed items may cause a decrease in item discrimination, a hypothesis we formally test with our RD model. Furthermore, the slope of item position appears linear on both sides of the cutoff, suggesting that a linear RD specification is appropriate. We provide additional descriptive analyses such as exploratory factor analysis scree plots, distributions of the latent trait estimates, and item characteristic curves for each survey in our supplement.

\subsection{Models}

\begin{figure}
    \centering
    \includegraphics[width=.75\linewidth]{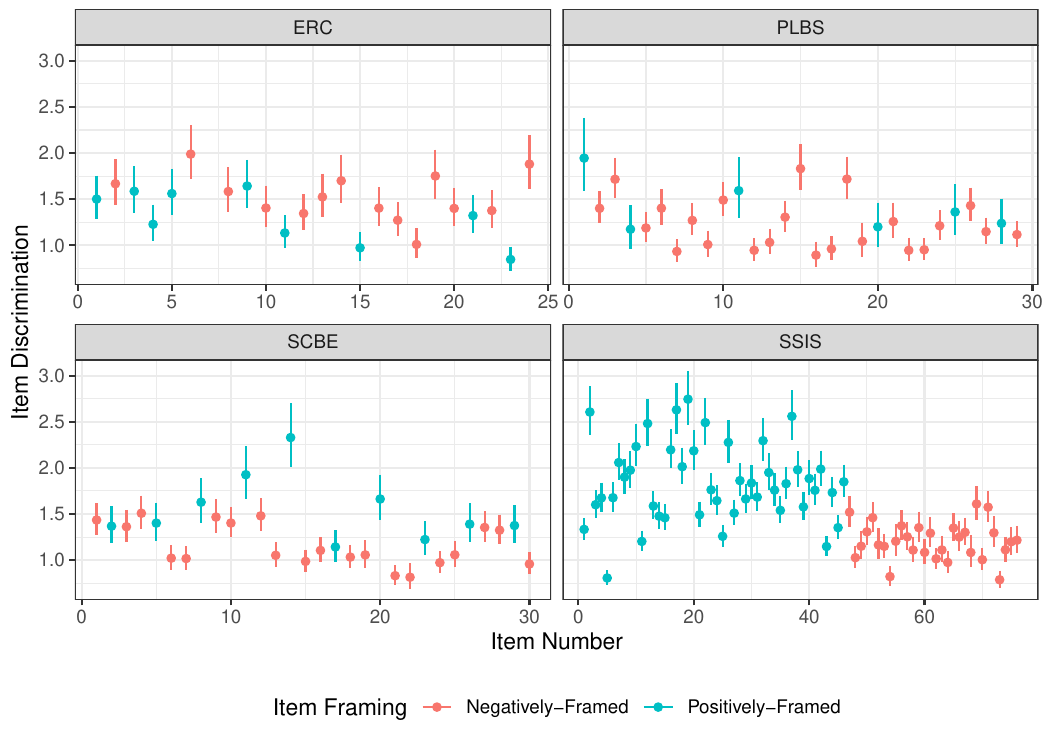}
    \caption{Posterior predictions of $a_i$ by survey}
    \label{fig:blups}
                 \justify \footnotesize
Notes: ERC = emotion regulation checklist; PLBS = preschool learning behavior scale; SCBE = social competence and behavior evaluation; SSIS = social skills improvement system. The bars show 95\% credible intervals. Points are color-coded by whether the item wording is positively or negatively framed.
\end{figure}

Table \ref{table:models} provides the results of the fitted models. The models confirm the substantial variability in $a_i$ observed in Figure \ref{fig:a_density}, with residual SDs of $\ln(a_i)$ of approximately .22 after accounting for differences in discrimination by negative framing. Across three of the four surveys, we observe a strong positive relationship between whether an item is negatively-framed and its location such that negatively-framed items tend to receive higher ratings, with $\widehat\beta_1>1$ for the ERC, SCBE, and SSIS. Because these items were reverse-coded, the large coefficient means that respondents tend to provide lower ratings on the original items. Simultaneously, negatively-framed items were substantially less discriminating than positively-framed items on two of the four surveys, with $\widehat\gamma_1$ of -.26 (SE = .09) and -.40 (SE = .05) on the log scale for the SCBE and SSIS, respectively. These are large differences, representing a 23\% and 33\% decrease in discriminating power, or 1.3 SDs and 1.8 SDs when indexed to $\sigma_a$ in a Cohen's \textit{d}-type metric. The effects of negative framing are less stark on the ERC and PLBS as the 95\% credible intervals include 0. Figure \ref{fig:blups} shows posterior predictions of the item discriminations with 95\% credible intervals for each item.

\begin{table}[ht]
\caption{Results of Explanatory Item Response Models}
\label{table:models}
\centering
\input{regtable}
\justify \footnotesize
Notes: The table presents the results of the fitted EIRMs. ERC = emotion regulation checklist; PLBS = preschool learning behavior scale; SCBE = social competence and behavior evaluation; SSIS = social skills improvement system; RD = regression discontinuity. Standard errors are in parentheses. $\sigma_\theta$ is constrained to 1 for model identification. Item discriminations are modeled on a log scale. The cut points (intercepts) for each model are denoted with $\le k$, representing the probability of selecting a response value $k$ or lower, conditional on the covariates in the model.\textbf{ }(Due to the non-linear nature of the model, the intercepts and coefficients do not directly translate into log-odds; they must be multiplied by the $a_i$ parameters first. For example, on the SSIS, $\beta_1 = 1.39$ must be multiplied by $e^{\gamma_0 + \gamma_1} = 1.20$ to provide the difference in agreeability between the average positively- and negatively-framed item on the logit scale.)
\end{table}

The RD model applied to the SSIS data shows that negative framing plausibly causes lower item discrimination ($\gamma_1 = -.36$), although the estimate is less precise than in the model for the difference in means. A randomization test in which we randomly assign the estimated $a_i$ to an item position and rerun the RD model 1,000 times provides an approximate p-value of .007 for this effect. The coefficients for the position of the item in the measure (i.e., $\gamma_2$ and $\gamma_2 + \gamma_3$ in Equation \ref{eq:rd}; not displayed in the table) are essentially 0 for both positively- and negatively-framed items, suggesting that it is not simply survey fatigue that predicts lower discrimination for the later items, but the switch to negative framing that causes a reduction in item discrimination.

The identifying assumption of the RD model could be compromised if, in addition to the introduction of negative framing, another simultaneous change occurred starting at item 47. For example, if item wording also became more complex and confused participants, this could also reduce item discrimination and confound the causal effect of the negative framing \citep{rush2016impact}. We view this as unlikely because a qualitative examination of the item text reveals that the only apparent difference beginning at item 47 is the negative framing. We test this hypothesis explicitly by applying the Flesch-Kincaid text complexity index to the SSIS item text and show that there is no change in readability when the items switch to negative framing. In fact, the negatively-framed items in the sample are slightly easier in readability than positively-framed items at the cutoff (effect size $ = -1.36$ grade levels, SE = 2.05, p = .51). A scatter plot of readability on item position is presented in Appendix \ref{appendix:reability}. Furthermore, when we include readability as a covariate in the model, we find that readability predicts neither location nor discrimination and that the effect of negative framing conditional on readability is essentially unchanged from our main specification ($\gamma_1 = -.35, \text{SE} = .10$).

An alternative threat to identification is that careless respondents artificially increase the difference in discrimination between the positively- and negatively-framed items \citep{ulitzsch2022explanatory, schroeders2022detecting, kam2023constrained, jaso2022identification, liu2024impact}. This could occur if, for example, some subset of respondents selected the same response category for each item in the survey. In our supplement, we test this possibility by excluding respondents with low values of intra-individual response variability (i.e., the within-person SD of item scores) as a proxy for careless responding \citep{yentes2020search, dunn2018intra, fiske1955intra}, and we find that the results of the analysis are essentially unchanged. More specifically, we exclude the 15\% of respondents with the lowest intra-individual response variability and refit the model to the remaining 85\% of respondents. In this subsample, we observe an estimated effect of $\gamma_1 = -.33, \text{SE} = .11$, a finding nearly identical to that reported in Table \ref{table:models}. In sum, given that the effect is so large in magnitude and that alternative explanations appear implausible, we view the causal argument that the switch to negative framing is causing lower item discrimination as the most reasonable interpretation of the results.\footnote{While the RD results are suggestive of a negative framing effect, it is nonetheless possible that it is the \textit{switch} from positive to negative framing, rather than negative framing \textit{per se}, that is driving the observed results. In other words, had the order of items been reversed, with the negative framed items first, a switch to positive framing part way through the survey could yield the same decrease in item discrimination. Distinguishing between these two potential mechanisms is not possible in the present setting.}

\section{Discussion}

Explaining item parameters as a function of item characteristics has a long history but has mainly focused on models for item location; models for item discrimination have remained comparatively unexplored. In this study, we show how the explanatory item response model (EIRM) that includes random effects for item discrimination and allows for latent regressions for the discriminations can be extended to polytomous responses and can answer substantive research questions about how negative framing may affect an item's discriminating power. 

Results from four social-emotional learning surveys applied to 1000 preschool children show some evidence that negatively-framed items are systematically less discriminating, with two of the four surveys showing substantially lower discrimination for negatively-framed items. Furthermore, a plausibly causal estimate from the regression discontinuity analysis shows that the sudden change to negatively-framed items partway through the survey causes lower item discrimination in that survey. A replication of the descriptive analyses using the followup survey data in Appendix \ref{appendix:t2} shows nearly identical results to the main analysis, providing further support for our findings.

Notably, two of the assessments in our sample (the PLBS and the ERC) do not show strong evidence of differences in item discrimination by negative framing. Indeed, the point estimate for the ERC is positive, though the credible interval includes 0. Thus, the effects of negative framing on item discrimination are likely to depend on the context, given that even in this single sample the effects are not consistent across surveys. We emphasize that the exploration of negative framing effects in this study is intended to be illustrative of the proposed method and provides a methodological framework for explicitly testing for such effects rather than serving as definitive evidence on the effects of negative framing on item discrimination more broadly. Larger-scale analyses across a range of samples and measures are needed to determine how consistent negative framing effects are and what sample, measure, or rater characteristics they might depend on, building on the extensive prior work in this area cited in Section \ref{intro}.

Our analysis is illustrative of the type of insights we can gain from extending the EIRM to model item discriminations as a function of covariates. However, we emphasize that even a finding suggesting that some types of items may be more or less informative than others on average should not be the sole criterion for including or excluding them from a measure. Other considerations such as construct underrepresentation, addressing potential acquiescence or agreement bias, or leveraging reverse-coded items in mixture models to detect careless behavior may be highly relevant depending on the assessment context \citep{jozsa2017reversed, ulitzsch2022explanatory, arias2023detecting, kam2023constrained}.

As a final note, we have contextualized our empirical exploration in terms of negative item framing effects because positively-framed items are generally the default against which other items are compared. In a certain sense, this vantage point is an arbitrary one. An equally valid interpretation of the results of the SSIS, for example, could be that positive framing \textit{inflates} item discrimination. As such, definition of the reference group and the interpretation of the results requires careful consideration and justification in a given research context.

While promising, several limitations temper the strength of our results. First, the Bayesian 2PL models explored here are considerably more computationally intensive than analogous frequentist 1PL models due to the MCMC estimation procedure, taking hours to fit rather than minutes or seconds, even with only 1000 respondents, and thus may present a challenge in large-scale data contexts. Second, while more flexible than other R packages, \texttt{brms} does not yet support a fully random effects specification of the GRM, either allowing fixed effects for all item thresholds, which precludes the use of item predictors in the model for item location, or fixed effects for the average item thresholds and uniform random shifts for the individual items, as explored here \citep[p. 36]{burkner2021brms}. If a more flexible approach is required, fitting the relevant model is possible directly in \texttt{Stan} \citep{carpenter2017stan} (for which \texttt{brms} is a wrapper) or \texttt{Mplus} \citep{Muthen2017}. As a sensitivity check, we refit our models with fixed threshold parameters for each item-category combination (i.e., allowing category distances to vary across items) with random effects for discrimination to better match the traditional GRM parameterization and include the results in our supplement. Overall, the magnitude and direction of the coefficients for negative framing are unchanged for the SSIS, effects remain non-significant for the PLBS. The effect for SCBE is no longer significant and the coefficient on the ERC remains positive but is now statistically significant. The standard deviation of $\ln(a_i)$, $\sigma_a$, is higher across all models. Thus, the results are somewhat sensitive to the choice of the IRT model.

Several extensions merit consideration as potential applications of the 2PL EIRM to evaluate differences in item discrimination:

\begin{enumerate}
    \item The model could easily be extended to include multiple variables for either richer description of how various features combine in predicting item discrimination, or non-uniform DIF analysis by including person covariates in the model for the discrimination parameter \citep{chalmers2018improving, breland2007investigating, narayanon1996identification}. For example, such an approach may shed light on how expert raters may provide more discriminating ratings than novices \citep{humphry2011role}. Such models are also closely related to moderated non-linear factor analysis approaches \citep{curran2014moderated}. We show example \texttt{brms} code to fit such models in Appendix \ref{appendix:code}.
    \item The models explored here impose linear relationships between the covariates and the item parameters. Recent developments that allow for smooth non-linear relationships between the covariates and the item parameters in an EIRM may allow for more nuanced explorations of item functioning \citep{cho2024modeling}.
    \item The application of causal inference methods to interventions designed to increase item discrimination, such as whether training raters increases the discrimination of their ratings, or how different wordings of the same item or different sets of distractors on multiple choice items can provide more discrimination in test design may lead to more informative and efficient measurement \citep{reeve2016psychometric}. 
    \item Research on modality effects on discrimination may provide insight into whether, for example, self-report versus external ratings of the same items, or different types of response options are more discriminating, which may have implications for cost and scale of assessments \citep{wang2021validating}.
    \item Models of item discrimination as a function of position in the test may help to reveal fatigue effects to inform optimal test length and could provide evidence regarding the need for counterbalancing or randomized item ordering \citep{nagy2018item, bulut2017structural, demirkol2022analyzing, debeer2013modeling, davis2005using, kingston1982effect, kingston1984item, hasselhorn2024effects, kanopka2024position}.  
    \item Leveraging covariates that vary at the person-item level such as response times or other types of process data and investigating how these are related to item discrimination could provide new insights into respondent behavior such as the speed-accuracy trade-off or detecting disengagement \citep{domingue2022speed, lu2023mixture, nagy2022multilevel, bolsinova2017modelling, ulitzsch2020hierarchical, ulitzsch2022response, koenig2023bayesian}. For instance, aberrantly fast response times may stem from disengaged behavior and may be associated with discriminations of essentially zero. Likewise, some types of cognitive assessment items may no longer discriminate between low- and high-ability respondents when respondents spend sufficient time on solving them. So far, such research questions were only possible to address with specialized software \citep{bolsinova2017modelling}.
    \item The models considered here use random intercepts for the item discrimination residuals; random slopes models that allow for the effects of person covariates on item discrimination to vary by item offer a promising extension for DIF analysis (or allowing the effects of item covariates to vary by person, see \cite{rijmen2002random}), based on similar work in 1PL causal inference models that allow unique treatment effects on each item by including random slopes of person-level treatment in the model \citep{gilbert2023jebs, gilbert2024ssri, gilbert2024disentangling}. 
    \item Conceptual work on the relationship between our model and network psychometrics \citep{isvoranu2022network, borsboom2022possible, epskamp2017generalized}, such as how discrimination parameters are related to the strength of network structures offer an area of potential synthesis between the latent variable approach explored here and alternative psychometric models \citep{christensen2021equivalency, marsman2018introduction, gilbert2024network}. For example, our analyses of the effects of negatively-framed items complement recent network psychometric findings illustrating that negatively-framed items function differently in psychometric network structures \citep{bulut2024psychometric}.
     \item Our approach can easily be integrated into factor analytic and SEM frameworks to facilitate applications to continuous response data. When used to study item wording effects, it may complement existing approaches such as bi-factor models  \citep{gu2015impact} commonly used to this end.
\end{enumerate}

\noindent In sum, extending the EIRM to 2PL settings, polytomous responses, and latent regression models of item discrimination parameters provides a useful opportunity to answer substantively interesting research questions that were difficult to explicitly address in prior frameworks. By applying these models, researchers in the psychological and behavioral sciences can gain important insights into why some items are more discriminating than others and can leverage this information to design more informative measures.

\newpage

\section{Declarations}

\textbf{Funding}: This research was supported in part by the Jacobs Foundation (Author 2).

\noindent \textbf{Conflicts of Interest}: The authors report no conflicts of interest.

\noindent \textbf{Ethics approval}: Not applicable.

\noindent \textbf{Consent to participate}: Not applicable.

\noindent \textbf{Consent for publication}: Not applicable.

\noindent \textbf{Availability of Data and Materials}: The original data set and survey questionnaires are publicly available at the following URL: \url{https://ldbase.org/datasets/38d4a723-c167-4908-a250-2cf29a4ff49b}. For convenience, we also include the raw datasets and codebook in our replication materials.

\noindent \textbf{Code Availability}: Our code, analysis output, and supplemental materials are available at the following URL: \url{https://researchbox.org/3008&PEER_REVIEW_passcode=TSPWAZ}.

\noindent \textbf{Author Contributions}:

\noindent Conceptualization: Author 1

\noindent Methodology: Author 1, Author 2, Author 4

\noindent Software: Author 1

\noindent Formal Analysis: Author 1

\noindent Writing---original draft preparation: Author 1

\noindent Writing---review and editing: Author 1, Author 2, Author 3, Author 4

\printbibliography

\newpage

\appendix
\section*{Appendix}

\renewcommand{\thesubsection}{\Alph{subsection}}

\subsection{Additional Notes on IRT Software} \label{appendix:software}

R provides a wide range of software for fitting IRT models. We review the most common packages here and note their limitations with respect to the 2PL EIRMs under investigation in this study.

\begin{enumerate}
    \item \texttt{mirt} \citep{chalmers2012mirt} is dedicated to IRT modeling and allows for a wide variety of models to be fit, but does not support random effects for item discrimination parameters.
    \item The SEM software \texttt{lavaan} \citep{rosseel2012lavaan} does not support random locations, random factor loadings, or the logistic link functions that are most common in IRT applications.
    \item  \texttt{flirt} \citep{jeon2016modular} supports 2PL EIRMs but does not provide standard errors for differences in item discrimination parameters \citep{min2018understanding}.
    \item  \texttt{PLmixed} \citep{rockwood2019estimating} and \texttt{galamm} \citep{sorensen2024multilevel} do not directly support polytomous responses, but ordinal data can be reformatted in such a way so that polytomous IRT models such as the rating scale model and the partial credit model can be estimated using the computational machinery of binary logistic regression \citep{bulut2021estimating, gilbert2024ssri}. Neither package supports random effects for item discriminations (Ø. Sørensen, personal communication, August 14, 2024).
    \item  \texttt{clmm} from the \texttt{ordinal} package \citep{christensen2015analysis} is designed for ordinal responses but does not accommodate factor loadings.
    \item  \texttt{GLMMadaptive} \citep{rizopoulos2022package} does not allow for crossed effects.
    \item  \texttt{hIRT} \citep{zhou2019hierarchical} allows for both ordinal and dichotomous item responses, but only accommodates fixed items and person predictors.
\end{enumerate}

\clearpage

\subsection{Item Text and Coding from Each Survey} \label{appendix:item_text}

\setcounter{table}{0}
\setcounter{figure}{0}
\renewcommand{\thetable}{B.\arabic{table}}
\renewcommand{\thefigure}{B.\arabic{figure}}

\input{app_item_tab}

\clearpage

\subsection{Sample \texttt{brms} code} \label{appendix:code}

The code below shows how to fit our basic ordinal model and some simple extensions in \texttt{brms}, where \texttt{resp} is a positive integer representing the item response, \texttt{neg} is an indicator variable for whether the item is negatively framed, \texttt{id} is a person identifier, and \texttt{item} is an item identifier. \texttt{disc} is a convenience parameter in \texttt{brms} that allows for easy specification of a model for item discrimination. The \texttt{|i|} syntax allows for a correlation between the location and discrimination random effects. The \texttt{cores} argument allows for parallel processing across multiple cores of the CPU. Note that for the priors on standard deviation parameters, \texttt{normal} indicates a half-normal distribution. See \textcite{burkner2021brms} for a full tutorial on fitting IRT models with \texttt{brms}, including dichotomous 1PL, 2PL, and 3PL models and their ordinal extensions.

\begin{singlespace}
\begin{verbatim}
# declare model
mod <- bf(
  resp ~ 1 + neg + (1|id) + (1|i|item),
  disc ~ 1 + neg + (1|i|item)
)

# set weakly informative priors
prior <- 
  # SD of item locations
  prior("normal(0, 1)", class = "sd", group = "item") +
  # fix SD of theta to be 1 for identification
  prior("constant(1)", class = "sd", group = "id") +
  # SD of discriminations
  prior("normal(0, .5)", class = "sd", group = "item", dpar = "disc") +
  # coefficients for location
  prior("normal(0, 1)", class = "b", coef = "neg") +
  # coefficients for discrimination
  prior("normal(0, .5)", class = "b", coef = "neg", dpar = "disc")

# fit model
fit <- brm(
   formula = mod,
   data = dataset_name,
   family = brmsfamily("cumulative", "logit"),
   prior = prior,
   backend = "cmdstanr",
   chains = 4,
   iter = 2000,
   cores = 4,
   threads = threading(4),
   refresh = 10
   )

# code to fit fixed thresholds and random discriminations
mod <- bf(
  resp | thres(gr = item) ~ 1 + (1|id),
  disc ~ 1 + neg + (1|item)
)

# person predictor Xj can be included in the equation
# for the response, to represent differences in theta
# or in the discrimination, to represent non-uniform DIF:
# (assuming no location shift beyond the average effect on theta)
mod <- bf(
  resp ~ 1 + neg + Xj + (1|id) + (1|i|item),
  disc ~ 1 + neg + Xj + (1|i|item)
)

# we can allow for both location and discrimination shifts
# for all items as follows:
mod <- bf(
  resp ~ 1 + neg + Xj + (1|id) + (Xj|i|item),
  disc ~ 1 + neg + Xj + (Xj|i|item)
)

# random slopes can be included in the random effects terms
# for example, to allow differences in agreeability by framing
# to vary by person:
mod <- bf(
  resp ~ 1 + neg + (neg|id) + (1|i|item),
  disc ~ 1 + neg + (1|i|item)
)

# code to fit a dichotomous 2PL model
# here, we declare the discrimination explicitly
# rather than use the disc parameter
# and resp is a 0/1 variable
mod <- bf(
  resp ~ exp(logalpha)*eta,
  eta ~ 1 + neg + (1|id) + (1|i|item),
  logalpha ~ 1 + neg + (1|i|item),
  nl = TRUE
)

# when fitting the model, the argument below uses
# standard dichotomous logit rather than cumulative logit
family = brmsfamily("bernoulli", "logit")

\end{verbatim}
\end{singlespace}

\clearpage

\subsection{Directed Acyclic Graph of Equation \ref{eq:egrm}} \label{appendix:dag}

\setcounter{table}{0}
\setcounter{figure}{0}
\renewcommand{\thetable}{D.\arabic{table}}
\renewcommand{\thefigure}{D.\arabic{figure}}
\begin{figure}[ht]
    \centering
    \includegraphics[width=.75\linewidth]{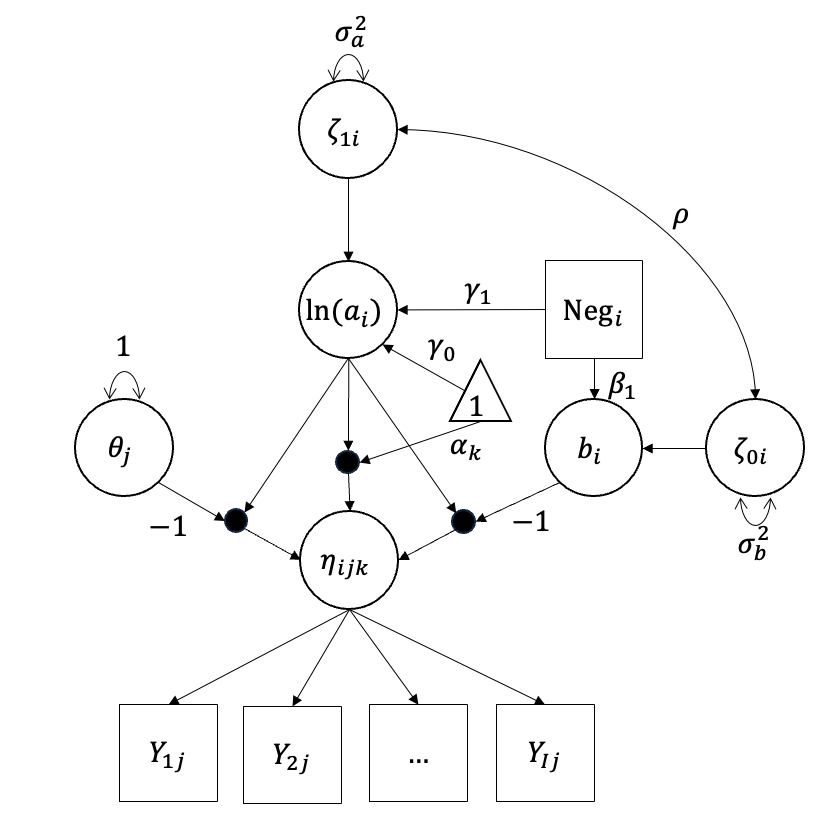}
    \caption{Directed Acyclic Graph for Equation \ref{eq:egrm}}
    \label{fig:dag}
                 \justify \footnotesize
Notes: Squares indicate observed variables, hollow circles indicate latent variables, and black circles indicate cross-product terms. $Y_{ij}$ are item responses, $I$ is the total number of items, and $\text{Neg}_i$ is the indicator for a negatively-framed item. Path coefficients are fixed at 1 unless otherwise indicated. The $\sigma^2$ terms represent residual variances.
\end{figure}

\clearpage
\subsection{Replication of Descriptive Analyses on Followup Survey Data} \label{appendix:t2}

\setcounter{table}{0}
\setcounter{figure}{0}
\renewcommand{\thetable}{E.\arabic{table}}
\renewcommand{\thefigure}{E.\arabic{figure}}

The graphs below replicate the descriptive analyses on the followup survey data in a sample of 470 children. The results are quite consistent across time points.

\begin{figure}[ht]
    \centering
    \includegraphics[width=.75\linewidth]{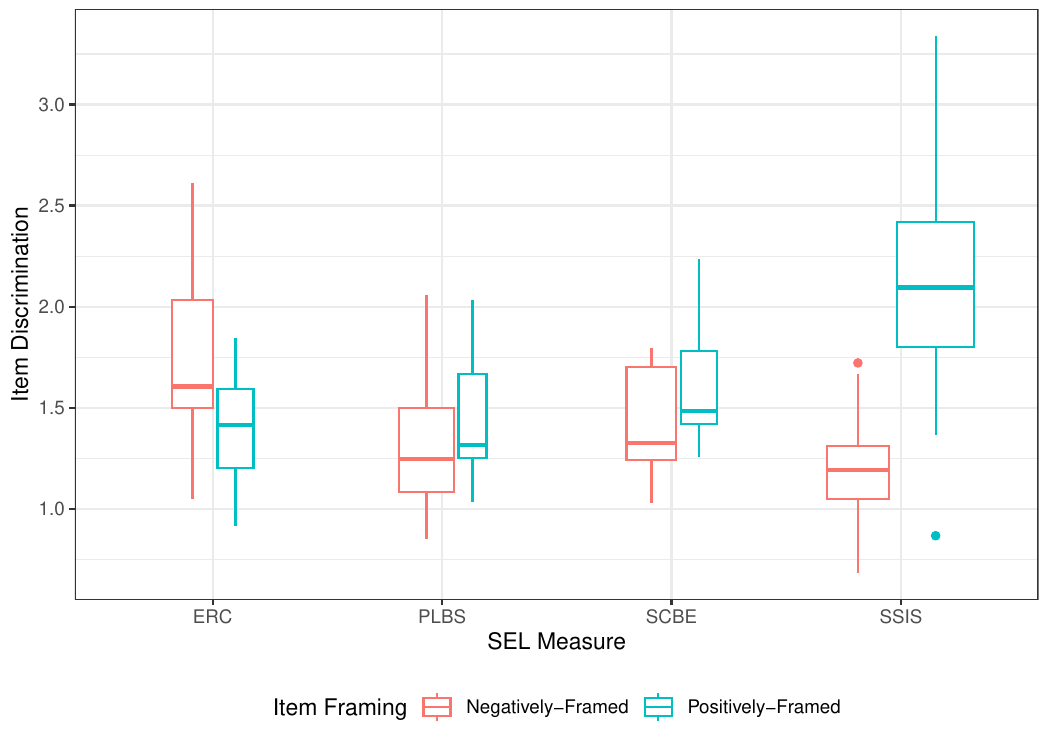}
    \caption{Boxplot of $a$ parameter estimates from each survey at Time 2}
                 \justify \footnotesize
Notes: ERC = emotion regulation checklist; PLBS = preschool learning behavior scale; SCBE = social competence and behavior evaluation; SSIS = social skills improvement system. The width of each box is proportional to the number of items. Boxes are color-coded by whether the item wording is positively or negatively framed.
\end{figure}

\begin{figure}[ht]
    \centering
    \includegraphics[width=.75\linewidth]{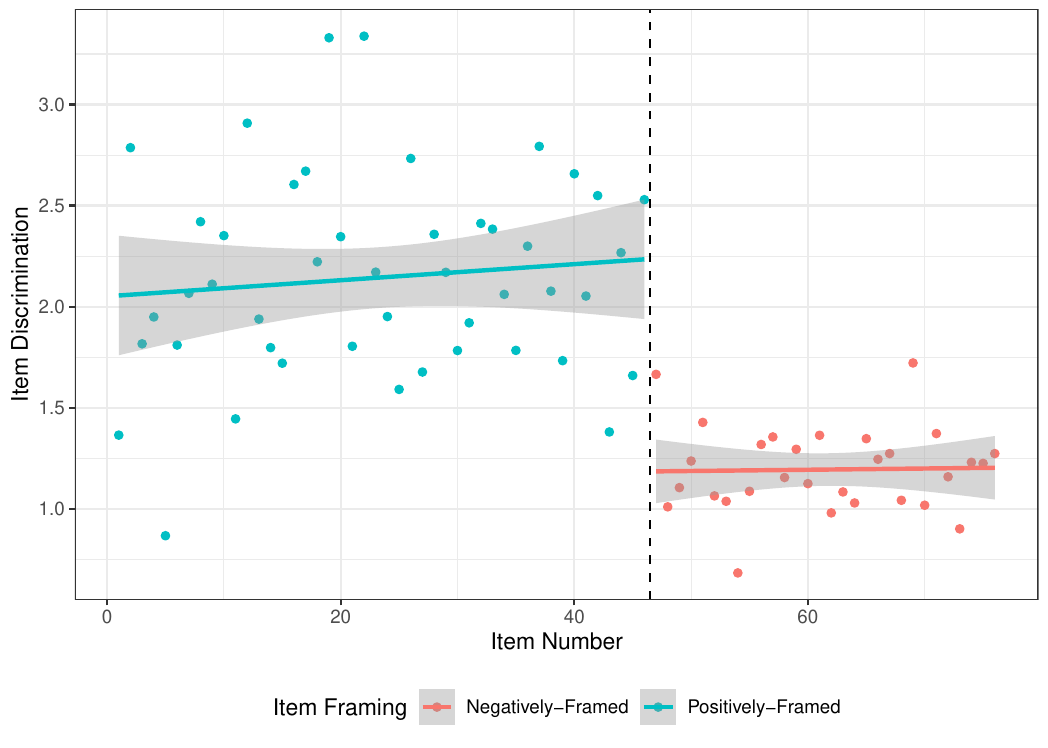}
    \caption{$a$ parameter estimates by item number for the SSIS at Time 2} 
                 \justify \footnotesize
Notes: SSIS = social skills improvement system. The lines and shaded areas represent OLS predictions and 95\% confidence intervals, respectively, fit separately to each item group. The points and lines are color-coded by whether the item wording is positively or negatively framed.
\end{figure}

\clearpage
\subsection{Readability Scatter Plot} \label{appendix:reability}

\setcounter{table}{0}
\setcounter{figure}{0}
\renewcommand{\thetable}{F.\arabic{table}}
\renewcommand{\thefigure}{F.\arabic{figure}}
\begin{figure}[ht]
    \centering
    \includegraphics[width=.75\linewidth]{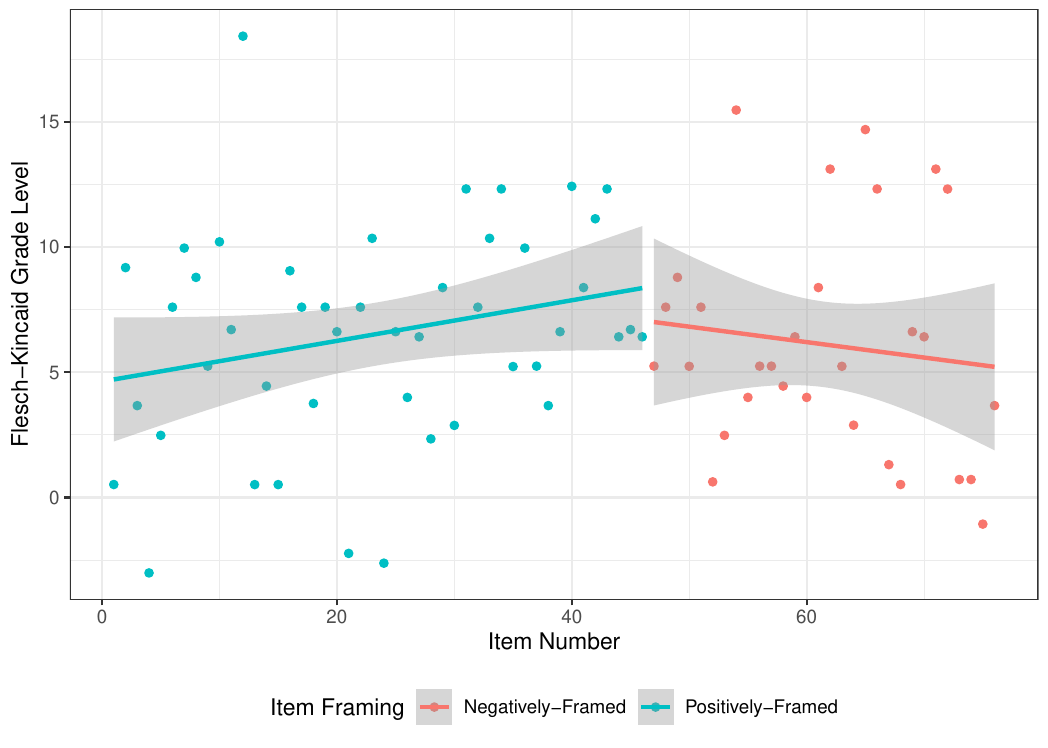}
    \caption{Flesch-Kincaid Grade Level readability by item number for the SSIS}
    \label{fig:rd_readability}
                 \justify \footnotesize
Notes: SSIS = social skills improvement system. Lower values indicate simpler text. The lines and shaded areas represent OLS predictions and 95\% confidence intervals, respectively, fit separately to each item group. The points and lines are color-coded by whether the item wording is positively or negatively framed.
\end{figure}

\end{document}

%% file: tab1.tex
\scalebox{.8}{

    \centering
    \begin{tabular}{ p{2in} p{1.5in} p{1.7in} p{0.2in} p{1.7in} p{0.2in}}
    \toprule
         \multirow{2}{*}{Measure} & \multirow{2}{*}{Rating Scale}  &  \multicolumn{2}{c}{Positive Items} &  \multicolumn{2}{c}{Negative Items} \\
         \cline{3-6}
         & & Example & N & Example & N \\
         \hline
         Emotion Regulation Checklist (ERC)& Never, Sometimes, Often, Almost Always (1-4)& Is able to delay gratification& 9&  Exhibits wide mood swings & 14\\ 
         \\
         Preschool Learning Behaviors Scale (PLBS)&   Does not apply, sometimes applies, most often applies (0-2) &  Pays attention to what you say& 6 &   Is reluctant to tackle a new activity& 23\\ 
         \\
         Social Competence and Behavior Evaluation (SCBE)&   Never, sometimes, often, always (1-6)&  Works easily in groups& 10 & Irritable, gets mad easily& 20\\
         \\
         Social Skills Improvement System (SSIS)&   Never, seldom, often, almost always (0-3) &  Says ``please''& 46 & Bullies others&30\\
         \bottomrule
    \end{tabular}

}

%% file: regtable.tex
\scalebox{.8}{
\centering
\begin{tabular}{llllll}
  \hline
  Estimate& ERC & PLBS & SCBE & SSIS & SSIS RD \\ 
  \hline
 $\le$1& -2.196 (0.19) & -2.526 (0.28) & -2.263 (0.24) & -2.217 (0.11) & -2.234 (0.11) \\ 
   $\le$2& -0.645 (0.18) & -0.385 (0.27) & -1.146 (0.24) & -0.71 (0.1) & -0.727 (0.1) \\ 
   $\le$3& 0.831 (0.18) &  & -0.404 (0.23) & 0.878 (0.11) & 0.863 (0.1) \\ 
   $\le$4&  &  & 0.41 (0.23) &  &  \\ 
   $\le$5&  &  & 1.507 (0.23) &  &  \\ 
   Mean ln(a) ($\gamma_0$) & 0.248 (0.08) & 0.332 (0.1) & 0.401 (0.07) & 0.579 (0.04) & 0.579 (0.07) \\ 
   b: Negatively-Framed ($\beta_1$) & 1.095 (0.24) & -0.084 (0.31) & 1.708 (0.29) & 1.385 (0.16) & 1.364 (0.15) \\ 
   a: Negatively-Framed ($\gamma_1$) & 0.155 (0.1) & -0.15 (0.1) & -0.261 (0.09) & -0.401 (0.05) & -0.361 (0.1) \\ 
   \hline
   $\sigma_b$ & 0.528 (0.1) & 0.694 (0.1) & 0.716 (0.1) & 0.676 (0.06) & 0.681 (0.06) \\ 
   $\sigma_a$ & 0.222 (0.04) & 0.23 (0.04) & 0.217 (0.03) & 0.226 (0.02) & 0.229 (0.02) \\ 
   $\sigma_\theta$ & 1& 1& 1& 1& 1\\ 
   $\rho_{ab}$& -0.003 (0.21) & -0.386 (0.17) & -0.332 (0.17) & 0.198 (0.11) & 0.206 (0.12) \\ 
   \hline
\end{tabular}

}

%% file: app_item_tab.tex
\singlespacing
\begin{footnotesize}
\begin{longtable}{>{\raggedright\arraybackslash}p{1.5cm}  
>{\raggedright\arraybackslash}p{1.5cm}  
                  >{\raggedright\arraybackslash}p{9cm} 
                  >{\raggedright\arraybackslash}p{1.5cm}
                  }

\caption{Item text for four SEL surveys} \label{tab:item_text}
\\

\centering
    Survey & Item Number & Item Text & Framing \\ 
    \toprule
    ERC & 1 & Is a cheerful child. & Positive \\
    ERC & 2 & Exhibits wide mood swings (child’s emotional state is difficult to anticipate because s/he moves quickly from positive to negative moods). & Negative \\
    ERC & 3 & Responds positively to neutral or friendly approaches by adults. & Positive \\
    ERC & 4 & Transitions well from one activity to another; does not become anxious, angry, distressed or overly excited when moving from one activity to another. & Positive \\
    ERC & 5 & Can recover quickly from episodes of upset or distress (e.g., does not pout or remain sullen, anxious or sad after emotionally distressing events). & Positive \\
    ERC & 6 & Is easily frustrated. & Negative \\
    ERC & 7* & Responds positively to neutral or friendly approaches by peers. & Positive \\
    ERC & 8 & Is prone to angry outbursts / tantrums easily & Negative \\
    ERC & 9 & Is able to delay gratification (wait for good things). & Positive \\
    ERC & 10 & Takes pleasure in the distress of others (e.g., laughs when another person gets hurt or punished; enjoys teasing others). & Negative \\
    ERC & 11 & Can modulate excitement in emotionally arousing situations (e.g., does not get ``carried away'' in high-energy situations, or overly excited in inappropriate contexts). & Positive \\
    ERC & 12 & Is whiny or clingy with adults. & Negative \\
    ERC & 13 & Is prone to disruptive outbursts of energy and exuberance. & Negative \\
    ERC & 14 & Responds angrily to limit-setting by adults. & Negative \\
    ERC & 15 & Can say when s/he is feeling sad, angry or mad, fearful or afraid. & Positive \\
    ERC & 16 & Seems sad or listless. & Negative \\
    ERC & 17 & Is overly exuberant when attempting to engage others in play. & Negative \\
    ERC & 18 & Displays flat affect (expression is vacant and inexpressive; child seems emotionally absent). & Negative \\
    ERC & 19 & Responds negatively to neutral or friendly approaches by peers (e.g., may speak in an angry tone of voice or respond fearfully). & Negative \\
    ERC & 20 & Is impulsive. & Negative \\
    ERC & 21 & Is empathetic towards others; shows concern when others are upset or distressed. & Positive \\
    ERC & 22 & Displays exuberance that others find intrusive or disruptive. & Negative \\
    ERC & 23 & Displays appropriate negative emotions (anger, fear, frustration, distress) in response to hostile, aggressive or intrusive acts by peers. & Positive \\
    ERC & 24 & Displays negative emotions when attempting to engage others in play. & Negative \\
    \hline
    PLBS & 1 & Pays attention to what you say. & Positive \\
    PLBS & 2 & Says task is too hard without making much effort to attempt it. & Negative \\
    PLBS & 3 & Is reluctant to tackle a new activity. & Negative \\
    PLBS & 4 & Sticks to an activity for as long as can be expected for a child of this age. & Positive \\
    PLBS & 5 & Adopts a don’t-care attitude to success or failure. & Negative \\
    PLBS & 6 & Seems to take refuge in helplessness. & Negative \\
    PLBS & 7 & Follows peculiar and inflexible procedures in tackling activities. & Negative \\
    PLBS & 8 & Shows little desire to please you. & Negative \\
    PLBS & 9 & Is unwilling to accept help even when an activity proves too difficult. & Negative \\
    PLBS & 10 & Acts without taking sufficient time to look at the problem or work out a solution. & Negative \\
    PLBS & 11 & Cooperates in group activities. & Positive \\
    PLBS & 12 & Bursts into tears when faced with a difficulty. & Negative \\
    PLBS & 13 & Has enterprising ideas which often don’t work out. & Negative \\
    PLBS & 14 & Is distracted too easily by what is going on in the room, or seeks distractions. & Negative \\
    PLBS & 15 & Cannot settle into an activity. & Negative \\
    PLBS & 16 & Gets aggressive or hostile when frustrated. & Negative \\
    PLBS & 17 & Is very hesitant in talking about his or her activity. & Negative \\
    PLBS & 18 & Shows little determination to complete an activity, gives up easily. & Negative \\
    PLBS & 19 & Uses headaches or other pains as a means of avoiding participation. & Negative \\
    PLBS & 20 & Is willing to be helped. & Positive \\
    PLBS & 21 & Is too lacking in energy to be interested in anything or to make much effort. & Negative \\
    PLBS & 22 & Relies on personal charm to get others to find solutions to the problems he or she meets. & Negative \\
    PLBS & 23 & Invents silly ways of doing things. & Negative \\
    PLBS & 24 & Doesn’t achieve anything constructive when in a mopey or sulky mood. & Negative \\
    PLBS & 25 & Shows a lively interest in the activities. & Positive \\
    PLBS & 26 & Tries hard but concentration soon fades and performance deteriorates. & Negative \\
    PLBS & 27 & Carries out tasks according to own ideas rather than in the accepted way. & Negative \\
    PLBS & 28 & Accepts new activities without fear or resistance. & Positive \\
    PLBS & 29 & Is dependent on adults for what to do, and takes few initiatives. & Negative \\
    \hline
    SCBE & 1 & Irritable, gets mad easily. & Negative \\
    SCBE & 2 & Negotiates solutions to conflicts with other children. & Negative \\
    SCBE & 3 & Remains apart, isolated from the group. & Negative \\
    SCBE & 4 & Easily frustrated. & Negative \\
    SCBE & 5 & Comforts or assists another child in difficulty. & Positive \\
    SCBE & 6 & Inactive, watches the other children play. & Negative \\
    SCBE & 7 & Defiant when reprimanded. & Negative \\
    SCBE & 8 & Takes other children and their point of view into account. & Positive \\
    SCBE & 9 & Sad, unhappy, or depressed. & Negative \\
    SCBE & 10 & Gets into conflict with other children. & Negative \\
    SCBE & 11 & Works easily in groups. & Positive \\
    SCBE & 12 & Inhibited or uneasy in the group. & Negative \\
    SCBE & 13 & Screams or yells easily. & Negative \\
    SCBE & 14 & Cooperates with other children in group activities. & Positive \\
    SCBE & 15 & Doesn’t talk or interact during group activities. & Negative \\
    SCBE & 16 & Gets angry when interrupted. & Negative \\
    SCBE & 17 & Takes pleasure in own accomplishments. & Positive \\
    SCBE & 18 & Timid, afraid (e.g., avoids new situations). & Negative \\
    SCBE & 19 & Hits, bites or kicks other children. & Negative \\
    SCBE & 20 & Accepts compromises when reasons are given. & Positive \\
    SCBE & 21 & Goes unnoticed in a group. & Negative \\
    SCBE & 22 & Hits teacher or destroys things when angry with teacher. & Negative \\
    SCBE & 23 & Attentive towards younger children. & Positive \\
    SCBE & 24 & Worries. & Negative \\
    SCBE & 25 & Forces other children to do things they don’t want to do. & Negative \\
    SCBE & 26 & Takes care of toys. & Positive \\
    SCBE & 27 & Tired. & Negative \\
    SCBE & 28 & Opposes the teacher’s suggestions. & Negative \\
    SCBE & 29 & Helps with everyday tasks (e.g., distributes snacks). & Positive \\
    SCBE & 30 & Maintains neutral facial expression (doesn’t smile or laugh). & Negative \\
    \hline
    SSIS & 1 & Asks for help from adults. & Positive \\
    SSIS & 2 & Follows your directions. & Positive \\
    SSIS & 3 & Tries to comfort others. & Positive \\
    SSIS & 4 & Says ``please.'' & Positive \\
    SSIS & 5 & Questions rules that may be unfair. & Positive \\
    SSIS & 6 & Is well-behaved when unsupervised. & Positive \\
    SSIS & 7 & Completes tasks without bothering others. & Positive \\
    SSIS & 8 & Forgives others. & Positive \\
    SSIS & 9 & Makes friends easily. & Positive \\
    SSIS & 10 & Responds well when others start a conversation or activity. & Positive \\
    SSIS & 11 & Stands up for herself/himself when treated unfairly. & Positive \\
    SSIS & 12 & Participates appropriately in class. & Positive \\
    SSIS & 13 & Feels bad when others are sad. & Positive \\
    SSIS & 14 & Speaks in appropriate tone of voice. & Positive \\
    SSIS & 15 & Says when there is a problem. & Positive \\
    SSIS & 16 & Takes responsibility for her/his own actions. & Positive \\
    SSIS & 17 & Pays attention to your instructions. & Positive \\
    SSIS & 18 & Shows kindness to others when they are upset. & Positive \\
    SSIS & 19 & Interacts well with other children. & Positive \\
    SSIS & 20 & Takes turns in conversations. & Positive \\
    SSIS & 21 & Stays calm when teased. & Positive \\
    SSIS & 22 & Acts responsibly when with others. & Positive \\
    SSIS & 23 & Joins activities that have already started. & Positive \\
    SSIS & 24 & Says ``thank you.'' & Positive \\
    SSIS & 25 & Expresses feelings when wronged. \\
    SSIS & 26 & Takes care when using other people’s things. & Positive \\
    SSIS & 27 & Ignores classmates when they are distracting. & Positive \\
    SSIS & 28 & Is nice to others when they are feeling bad. & Positive \\
    SSIS & 29 & Invites others to join in activities. & Positive \\
    SSIS & 30 & Makes eye contact when talking. & Positive \\
    SSIS & 31 & Takes criticism without getting upset. & Positive \\
    SSIS & 32 & Respects the property of others. & Positive \\
    SSIS & 33 & Participates in games or group activities. & Positive \\
    SSIS & 34 & Uses appropriate language when upset. & Positive \\
    SSIS & 35 & Stands up for others who are treated unfairly. & Positive \\
    SSIS & 36 & Resolves disagreements with you calmly. & Positive \\
    SSIS & 37 & Follows classroom rules. & Positive \\
    SSIS & 38 & Shows concern for others. & Positive \\
    SSIS & 39 & Starts conversations with peers. & Positive \\
    SSIS & 40 & Uses gestures or body appropriately with others. & Positive \\
    SSIS & 41 & Responds appropriately when pushed or hit. & Positive \\
    SSIS & 42 & Takes responsibility for part of a group activity. & Positive \\
    SSIS & 43 & Introduces himself/herself to others. & Positive \\
    SSIS & 44 & Makes a compromise during a conflict. & Positive \\
    SSIS & 45 & Says nice things about herself/himself without bragging. & Positive \\
    SSIS & 46 & Stays calm when disagreeing with others. & Positive \\
    SSIS & 47 & Acts without thinking. & Negative \\
    SSIS & 48 & Is preoccupied with object parts. & Negative \\
    SSIS & 49 & Bullies others. & Negative \\
    SSIS & 50 & Becomes upset when routines change. & Negative \\
    SSIS & 51 & Has difficulty waiting for turn. & Negative \\
    SSIS & 52 & Does things to make others feel scared. & Negative \\
    SSIS & 53 & Fidgets or moves around too much. & Negative \\
    SSIS & 54 & Has stereotyped motor behaviors. & Negative \\
    SSIS & 55 & Forces others to act against their will. & Negative \\
    SSIS & 56 & Withdraws from others. & Negative \\
    SSIS & 57 & Has temper tantrums. & Negative \\
    SSIS & 58 & Keeps others out of social circles. & Negative \\
    SSIS & 59 & Breaks into or stops group activities. & Negative \\
    SSIS & 60 & Repeats the same thing over and over. & Negative \\
    SSIS & 61 & Is aggressive towards people or objects. & Negative \\
    SSIS & 62 & Gets embarrassed easily. & Negative \\
    SSIS & 63 & Cheats in games or activities. & Negative \\
    SSIS & 64 & Acts lonely. & Negative \\
    SSIS & 65 & Is inattentive. & Negative \\
    SSIS & 66 & Has nonfunctional routines or rituals. & Negative \\
    SSIS & 67 & Fights with others. & Negative \\
    SSIS & 68 & Says bad things about self. & Negative \\
    SSIS & 69 & Disobeys rules or requests. & Negative \\
    SSIS & 70 & Has low energy or is lethargic. & Negative \\
    SSIS & 71 & Gets distracted easily. & Negative \\
    SSIS & 72 & Uses odd physical gestures in interactions. & Negative \\
    SSIS & 73 & Talks back to adults. & Negative \\
    SSIS & 74 & Acts sad or depressed. & Negative \\
    SSIS & 75 & Lies or does not tell the truth. & Negative \\
    SSIS & 76 & Acts anxious with others. & Negative \\
  \bottomrule

   \hline
\end{longtable}

\justify
Notes: ERC = emotion regulation checklist; PLBS = preschool learning behavior scale; SCBE = social competence and behavior evaluation; SSIS = social skills improvement system. *ERC 7 was skipped in error during the baseline administration.

\end{footnotesize}
\doublespacing